\documentclass[%
 reprint,
 amsmath,amssymb,
 aps,
]{revtex4-1}

\usepackage{graphicx}% Include figure files
\usepackage{dcolumn}% Align table columns on decimal point
\usepackage{bm}% bold math
\usepackage{tabularx}
\usepackage{longtable}
\usepackage{threeparttable}
\usepackage{multirow}
\usepackage{rotating}
\usepackage{float}
\usepackage[]{units}
\usepackage{float}

\newcommand{\Lower}[1]{\smash{\lower 1.5ex \hbox{#1}}}
\begin{document}

\preprint{APS/ThOandThS}

\title{Electron correlation effects of the ThO and ThS molecules in the spinor basis. A relativistic coupled cluster study of ground and excited states properties.}
\author{Pawe{\l} Tecmer}
\affiliation{%
Institute of Physics, Faculty of Physics, Astronomy and Informatics, Nicolaus Copernicus University in Torun, Grudziadzka 5, 87-100 Torun, Poland,
\email{ptecmer@fizyka.umk.pl}
}
\author{Cristina E. Gonz\'alez-Espinoza}
\affiliation{%
 Department of Chemistry and Chemical Biology, McMaster University, Hamilton, 1280 Main Street West, L8S 4M1, Canada}

\date{\today}% It is always \today, today,

\begin{abstract}
We present a comprehensive relativistic coupled cluster study of the electronic structures of the ThO and ThS molecules in the spinor basis.
Specifically, we use the single-reference coupled cluster and the multi-reference Fock Space Coupled Cluster (FSCC) methods to model their ground and electronically-excited states.
Two variants of the FSCC method have been investigated: (a) one where the electronic spectrum is obtained from sector (1,1) of the Fock space, and (b) another where the excited states come from the doubly attached electronic states to the doubly charged systems (ThO$^{2+}$ and ThS$^{2+}$), that is, from sector (0,2) of the Fock space.
Our study provides a reliable set of spectroscopic parameters such as bond lengths, excitation energies, and vibrational frequencies, as well as a detailed analysis of the electron correlation effects in the ThO and ThS molecules.
Finally, we examine the first ionization potential and electron affinity of the above mentioned molecules.
\end{abstract}

\maketitle
%%%MAIN TEXT%%%%
%%%%%%%%%%%%%%%%%%%%%%%%%%%%%%%%%%%%%%%%%%%%%%%%%%%%%%%%%%%%%%%%%%%%%%%%%%%%%%%%%%%%%%%%%%%%%%%%%%%%%%%%%%%%%%%%%%%%%%%%%%
%%%%%%%%%%%%%%%%%%%%%%%%%%%%%%%%%%%%%%%%%%%%%%%%%%%%%%%%%%%%%%%%%%%%%%%%%%%%%%%%%%%%%%%%%%%%%%%%%%%%%%%%%%%%%%%%%%%%%%%%%%
\section{Introduction}
%%%%%%%%%%%%%%%%%%%%%%%%%%%%%%%%%%%%%%%%%%%%%%%%%%%%%%%%%%%%%%%%%%%%%%%%%%%%%%%%%%%%%%%%%%%%%%%%%%%%%%%%%%%%%%%%%%%%%%%%%%
%%%%%%%%%%%%%%%%%%%%%%%%%%%%%%%%%%%%%%%%%%%%%%%%%%%%%%%%%%%%%%%%%%%%%%%%%%%%%%%%%%%%%%%%%%%%%%%%%%%%%%%%%%%%%%%%%%%%%%%%%%
Coupled cluster theory can be considered as one of the most accurate approaches for practical \textit{ab initio} calculations of ground and electronically excited-state properties of atoms and molecules~\cite{Coester_1958,Cizek_jcp_1966,Cizek_Paldus_1971,Paldus_Cizek_Shavitt_1972,Bartlett_rev_1981,Helgaker_book,Shavitt_book,bartlett_2007}. 
This is particularly true for systems prevailed by so-called dynamic electron correlation effects, where the electronic wavefunction is well described by a single electron 
configuration. 
Ground state electronic structure properties can then be efficiently determined from the Coupled Cluster Singles and Doubles (CCSD) method or the CCSD(T) approach, which additionally includes a perturbative triples correction.  
While these methods are capable of reproducing highly accurate experimental data~\cite{heat-gauss,Borkowski2013}, a reliable and efficient description of electronic excited states remains more challenging. 
This led to the development of numerous methods with a CCSD reference state and where a spectrum of electronic states is obtained in a single calculation.  
Examples are the Equation of Motion Coupled Cluster Singles and Doubles (EOM-CCSD) method~\cite{eom-rowe-1968,eom-bartlett-1984,bartlett-eom,jankowski-eom}, including its spin-flip~\cite{spin-flip-EOM,spin-flip-EOM-theory,SOC-EOM}, completely renormalized~\cite{CR-EOMCCSD,EOMCCSd,unrestricted_CR-EOMCCSD,Kowalski_fused_porphyrine} and simplified variants~\cite{GWALTNEY1996189,leom-bartlett-2015,Kasia-eom-jcp-2016,Kasia-eom-jcp-2016-erratum}, as well as the Fock Space Coupled Cluster (FSCC) group of methods~\cite{meissner-musial-chapter,monika_mrcc,meissner-fscc-2012}. 
The major advantages of the FSCCSD method over the EOM-CCSD approach are the size extensivity of electronically-excited states and correct description of charge transfer excitations~\cite{FSCC_20_sector}. 

The Intermediate Hamiltonian (IH) formulation of the FSCCSD method represents a versatile tool to model excited states with multi-reference character~\cite{meissner-musial-chapter,monika_mrcc}.
Moreover, if coupled with a proper relativistic Hamiltonian, the IHFSCCSD method allows for a reliable description of excited states of heavy-element compounds~\cite{Timo_overview,U_Ivan}. 
Therefore, a relativistic formulation of this scheme in the spinor basis (comprised of Kramers pairs)~\cite{IHFSCC_3,IHFSCC_2} often serves as a reference method for quantum chemical modelling of the electronic spectra of small actinide species~\cite{CUO_Ivan,fscc-npo2,real09,pawel3,tecmer:pereira:ekstrom:visscher,tecmer_ihfsccsd,Cs2UO2Cl4-Gomes,ThF+2015}. 

Prototypical di- and tri-atomic molecules containing one actinide element are valuable models to examine bonding mechanisms and electronic structures of larger realistic actinide compounds~\cite{Cs2UO2Cl4-Gomes,pawel_saldien,pawel_PCCP2015,Boguslawski2017}. 
Such model compounds are, for instance, instructive to elucidate the participation of the 5$f$ and 6$d$ actinide orbitals in chemical bonding and their influence on molecular properties~\cite{Actinides_bible}. 

Actinide oxides and their derivatives~\cite{Kovacs2015} are one of the most explored small actinide compounds, by both, experimental and quantum chemical techniques.
The first and simplest representative of this group, the thorium monoxide (ThO) molecule, has been pointed out as a candidate in the search of the electron electric dipole moment (eEDM)~\cite{heaven:barker:antonov,wang:le:steimle:heaven,ACME_science}. 
As a result, both experimental and theoretical groups set sights on the reliable description of the ground and electronically excited states of the ThO molecule. 
Such information is crucial to estimate the lower bound for the permanent electric dipole moments in the $\text{X}^1\Sigma^+$ and $\text{H}^1 \Delta_1$ electronic states of ThO, which has been recently set to 8.7x$10^{-29} e$ cm$^{-1}$ by Baron~\textit{et al.}~\cite{ACME_science}. 
Additional experimental studies on ThO include gas phase microwave~\cite{exp_dewberry} and infrared measurements~\cite{exp_kushto,exp_andrews_jacs}, as well as high resolution photoelectron spectroscopy analysis~\cite{exp_edvinsson, exp_goncharov1, exp_goncharov2}. Theoretical examination covers multi-reference methods~\cite{Marian_tho_1988,theo_casscf_paulovic,PhysRevA.78.010502,edm_fleig_nayak2014,Skripnikov-ThO-2015,Skripnikov-ThO-2016,Denis-ThO-2016}, the single-reference coupled cluster approach~\cite{theo_dk3_paulovic,Buchachenko_2010,Titov-ThO}, and density functional theory calculations~\cite{ThS_liang_andrews_2002, Pereira-ths}.

Recently, Heaven and coworkers~\cite{heaven:barker:antonov,Sulfur_Heaven_2014} designated the ThS molecule as the new and potentially good candidate for the eEDM. 
Their preliminary \textit{ab initio} calculations confirm experimental findings~\cite{heaven:barker:antonov}.  
This motivates us to carefully examine the electronic structures of both ThO and ThS using state-of-the-art relativistic coupled cluster methods. 
We would like to stress that the goal of our work is not a direct determination of the lower bound for eEDM, but an in-depth examination of the electronic structures of ThO and ThS by pointing out similarities and differences between them.  
Our relativistic coupled cluster data including adiabatic excitation energies can further be used in the analysis of the lower bound for the permanent electric dipole moments. 
To the best of our knowledge, such a reliable theoretical study conducted for both ThO and ThS molecules has not been reported, yet. 
 
This article is organized as follows. A brief description of the IHFSCCSD method is presented in section~\ref{sec:fscc}. Computational details are presented in section~\ref{sec:compdetails}.  In section~\ref{sec:results}, we discuss our coupled cluster results for the ground, excited states, ionization potentials, and electron affinities of the ThO and ThS molecules. Finally, we conclude in section~\ref{sec:conclusions}.
%%%%%%%%%%%%%%%%%%%%%%%%%%%%%%%%%%%%%%%%%%%%%%%%%%%%%%%%%%%%%%%%%%%%%%%%%%%%%%%%%%%%%%%%%%%%%%%%%%%%%%%%%%%%%%%%%%%%%%%%%%
%%%%%%%%%%%%%%%%%%%%%%%%%%%%%%%%%%%%%%%%%%%%%%%%%%%%%%%%%%%%%%%%%%%%%%%%%%%%%%%%%%%%%%%%%%%%%%%%%%%%%%%%%%%%%%%%%%%%%%%%%%
%  THE FSCC method
%%%%%%%%%%%%%%%%%%%%%%%%%%%%%%%%%%%%%%%%%%%%%%%%%%%%%%%%%%%%%%%%%%%%%%%%%%%%%%%%%%%%%%%%%%%%%%%%%%%%%%%%%%%%%%%%%%%%%%%%%%
%%%%%%%%%%%%%%%%%%%%%%%%%%%%%%%%%%%%%%%%%%%%%%%%%%%%%%%%%%%%%%%%%%%%%%%%%%%%%%%%%%%%%%%%%%%%%%%%%%%%%%%%%%%%%%%%%%%%%%%%%%
\section{The Fock-space coupled cluster approach}\label{sec:fscc}
\begin{figure}[h!]
\begin{center}
\includegraphics[width=0.5\textwidth]{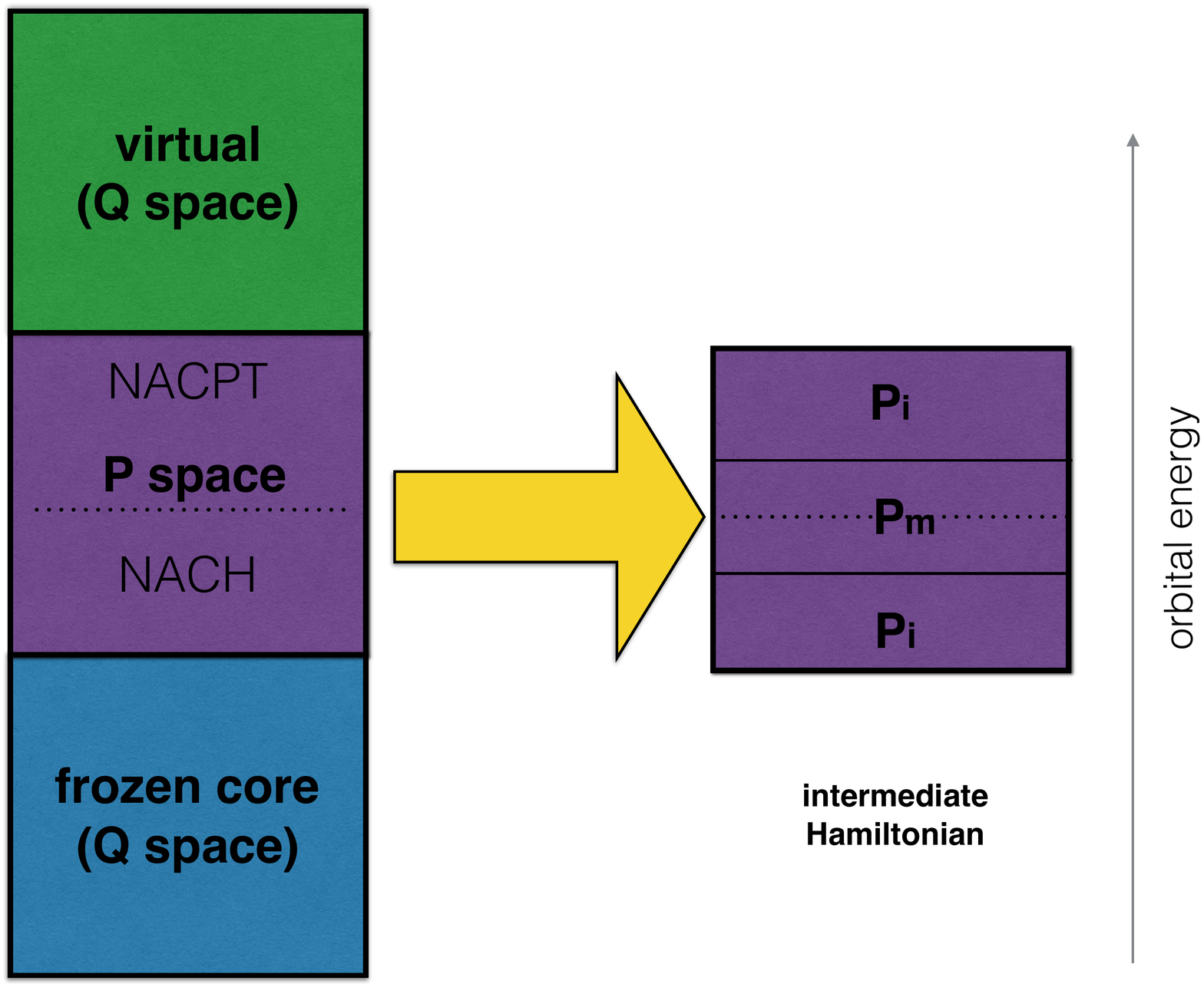}
\caption{Schematic representation of the IHFSCC model. The active space $\textbf{P}$ is the sum of the number of active holes (NACTH) and the number of active particles (NACTP). $\bf{P_i}$ is the intermediate Hamiltonian space, $\bf{P_m}$ is the model space, and $\bf{Q}$ is the auxiliary space of inactive orbitals. Only the excitations within the model space $\bf{P_m}$ have a physical meaning. }
\label{fig:scheme}
\end{center}
\end{figure}

The FSCC method is a state-universal multi-reference coupled cluster theory, which, as the name implies, operates in the Fock space.  
The basic idea of the FSCC method is to find an effective Hamiltonian in a low-dimensional model $P$ space, with eigenvalues approximating some desirable eigenvalues of the physical Hamiltonian ($H$). 
While the model or $P$ space contains the active valence orbitals directly involved in the electronic excitations, the complementary $Q$ space includes all the remaining orbitals. 
In this way, only a few eigenvalues out of the whole spectrum are calculated, and the expensive diagonalization of the $H$ Hamiltonian in the large configurational space is avoided. 
The use of a model space, however, might lead to intruder state problems, which are the source of divergencies encountered in certain basis sets or molecular geometries. 
To remedy this problem, the intermediate Hamiltonian formulation of the FSCC method has been proposed, which imposes a buffer space between the desired and undesired states.
This means that the model space ($P$) is further divided into a main model space ($P_m$) and an intermediate model space ($P_i$), which serves as a buffer between the $P_m$ and $Q$ spaces. A schematic representation of the IHFSCC model is presented in Figure~\ref{fig:scheme}. We should note that in our studies we used the IH scheme proposed by Kaldor and coworkers~\cite{IHFSCC_2}, but alternative techniques also exist~\cite{meissner-musial-chapter}. 
For more details about the IHFSCC method, we refer the reader to the literature~\cite{meissner-musial-chapter,monika_mrcc}. 
%%%%%%%%%%%%%%%%%%%%%%%%%%%%%%%%%%%%%%%%%%%%%%%%%%%%%%%%%%%%%%%%%%%%%%%%%%%%%%%%%%%%%%%%%%%%%%%%%%%%%%%%%%%%%%%%%%%%%%%%%%
%%%%%%%%%%%%%%%%%%%%%%%%%%%%%%%%%%%%%%%%%%%%%%%%%%%%%%%%%%%%%%%%%%%%%%%%%%%%%%%%%%%%%%%%%%%%%%%%%%%%%%%%%%%%%%%%%%%%%%%%%%
%  Figure 1
%%%%%%%%%%%%%%%%%%%%%%%%%%%%%%%%%%%%%%%%%%%%%%%%%%%%%%%%%%%%%%%%%%%%%%%%%%%%%%%%%%%%%%%%%%%%%%%%%%%%%%%%%%%%%%%%%%%%%%%%%%
%%%%%%%%%%%%%%%%%%%%%%%%%%%%%%%%%%%%%%%%%%%%%%%%%%%%%%%%%%%%%%%%%%%%%%%%%%%%%%%%%%%%%%%%%%%%%%%%%%%%%%%%%%%%%%%%%%%%%%%%%%
%%%%%%%%%%%%%%%%%%%%%%%%%%%%%%%%%%%%%%%%%%%%%%%%%%%%%%%%%%%%%%%%%%%%%%%%%%%%%%%%%%%%%%%%%%%%%%%%%%%%%%%%%%%%%%%%%%%%%%%%%%

An intrinsic feature of the FSCC approach is the partitioning into sectors ($k$,$l$) depending on the number of electrons removed from or attached to the reference state. 
In this way, the single electron excitation energies are obtained from sector ($1$,$1$), singly attached ones form sector ($0$,$1$), singly ionized ones from sector ($1$,$0$), doubly attached ones from sector ($0$,$2$), and doubly ionized ones from sector ($2$,$0$) of the Fock space. 
It is important to note that for arbitrary $k$ and $l$ sectors, subsystem embedding conditions require \textit{a priori} solution to all lower rank sectors~\cite{cluster-embedding-condition}. 
This means that before we calculate excitation energies from sector ($1$,$1$) of the Fock space, we first have to solve (and converge) equations for the ($0$,$0$), ($0$,$1$), and ($1$,$0$) sectors, respectively. 
Similarly, for excitation energies from sector ($0$,$2$) one has to find \textit{a priori} solutions to sectors ($0$,$0$) and ($0$,$1$), respectively.  

The FSCCSD(1,1) method cannot always be applied to calculate excited states of closed shell systems due to convergence difficulties originating from intruder states.  
This is particularly true when large basis sets are applied. 
For example, undesired (canonical) Rydberg-type orbitals with energies close to those in the model space might cause convergence difficulties~\cite{pawel3}.   
One way to overcome this problem is to use molecular orbitals from a charged system, which pushes such Rydberg-type orbitals energetically further away from the model space. 
Utilizing the FSCCSD(0,2) approach rises, however, the question about orbital relaxation effects and the reliability of obtained excitation energies computed with different Fock operators (with and without the presence of two electrons).   
%%%%%%%%%%%%%%%%%%%%%%%%%%%%%%%%%%%%%%%%%%%%%%%%%%%%%%%%%%%%%%%%%%%%%%%%%%%%%%%%%%%%%%%%%%%%%%%%%%%%%%%%%%%%%%%%%%%%%%%%%%
%%%%%%%%%%%%%%%%%%%%%%%%%%%%%%%%%%%%%%%%%%%%%%%%%%%%%%%%%%%%%%%%%%%%%%%%%%%%%%%%%%%%%%%%%%%%%%%%%%%%%%%%%%%%%%%%%%%%%%%%%%
%  Computational details
%%%%%%%%%%%%%%%%%%%%%%%%%%%%%%%%%%%%%%%%%%%%%%%%%%%%%%%%%%%%%%%%%%%%%%%%%%%%%%%%%%%%%%%%%%%%%%%%%%%%%%%%%%%%%%%%%%%%%%%%%%
%%%%%%%%%%%%%%%%%%%%%%%%%%%%%%%%%%%%%%%%%%%%%%%%%%%%%%%%%%%%%%%%%%%%%%%%%%%%%%%%%%%%%%%%%%%%%%%%%%%%%%%%%%%%%%%%%%%%%%%%%%
\section{Computational Details}\label{sec:compdetails}
All calculations have been carried out in the~\textsc{DIRAC14} relativistic software package~\cite{DIRAC14}. 
Through out this work, we used the (spin--orbit) Dirac--Coulomb Hamiltonian, in which the $\left(SS|SS \right )$ integrals were approximated by a point charge model~\cite{lvcorr}.  
In all the calculations we used a valence triple-$\zeta$ basis set of Dyall~\cite{dyall_b} if not stated otherwise. 
The correlated calculations used the so-called ``no-pair approximation'' where the projection operators remove any Slater determinant containing negative-energy orbitals from the Dirac--Coulomb Hamiltonian~\cite{Trond_rev_2012}.
In all calculations, the C$_{2v}^*$ double point group symmetry was employed~\cite{Dyall_book}. 

In the ground state CCSD and CCSD(T) calculations, we investigated various numbers of correlated electrons ranging from 28 to 50 and form 28 to 58 for ThO and ThS, respectively. 
We also studied the basis set effect on ground-state spectroscopic constants by extrapolation to the basis set limit.    
The basis set limit of the Dirac--Hartree--Fock energy was obtained by fitting an exponential function of the form~\cite{Helgaker1997}
        \begin{equation}
            E^{\mathrm{SCF}}({X}) = E^{\mathrm{SCF}}_{\infty} + a \exp(-b X)
        \end{equation}
to the Dirac--Hartree--Fock energies obtained in the cc-pVDZ, cc-pVTZ, and cc-pVQZ basis sets~\cite{dunning_b} for light elements (O,S) and the dyall.v2z, dyall.v3z, and dyall.v4z basis sets for thorium.~\cite{dyall_b} 
In the above equation $X$ indicates the cardinal number of the basis set (2 for D, 3 for T, etc.). 
For all correlated calculations, the basis set limit of the correlation energy was obtained by a two-point fit using the fit function
        \begin{equation}
            E^{\mathrm{corr}}({X}) = E^{\mathrm{corr}}_{\infty} + a X^{-3},
        \end{equation}
as suggested in refs.~\cite{Helgaker1997,Halkier1998a}.
In the above equation, $E^{\mathrm{corr}}({X})$ indicates the correlation energy of a given method defined as $E^{\mathrm{corr}}({X})= E^{\mathrm{tot}}({X})-E^{\mathrm{SCF}}({X})$.
The virtual spinors with energies above 30$E_{\text h}$ were deleted in the CC calculations. 

In our excited state calculations, we used two variants of the IHFSCC approach: one with the electronic spectra calculated from sector ($1$,$1$) of the Fock space (denoted as IHFSCC($1$,$1$)) and another with the electronic spectra obtained from the doubly attached electrons to the doubly charge species, that is, from sector (0,2) of the Fock space (denoted as IHFSCC(0,2)). 
In the CCSD reference state, we thus correlated 34 and 32 electrons for the IHFSCC(1,1) and IHFSCC(0,2) models, respectively. 
In all excited state calculations our main model space (P$_m$) was composed of the thorium 7$s$, 6$d$, and 7$p$ atomic spinors. 
The first ionization potentials and electron affinities were obtained from sectors (1,0) and (0,1) of the Fock space taking into account structural changes of ionized and electron attached species. 
More detailed information concerning the composition of each active space can be found in Tables S1 and S2 of the ESI\dag.  

The potential energy curves were obtained from a fith-order polynomial function fit to the single-point calculations in the range of 1.78--1.92~\AA~and 2.32--2.44~\AA~for the ThO and ThS molecules, respectively. 
Each fit was based on a single point calculation, displaced by 0.02~\AA~ around the equilibrium bond lengths.  
All the spectroscopic constants such as equilibrium bond lengths, harmonic vibrational frequencies, and force constants were calculated using the \textsc{twofit} program provided in DIRAC. 
In the calculation of our spectroscopic constants we have used masses of the most abundant isotopes, that is, 15.9949 for O, 31.9721 for S, and 232.0381 for Th.  

%%%%%%%%%%%%%%%%%%%%%%%%%%%%%%%%%%%%%%%%%%%%%%%%%%%%%%%%%%%%%%%%%%%%%%%%%%%%%%%%%%%%%%%%%%%%%%%%%%%%%%%%%%%%%%%%%%%%%%%%%%
%%%%%%%%%%%%%%%%%%%%%%%%%%%%%%%%%%%%%%%%%%%%%%%%%%%%%%%%%%%%%%%%%%%%%%%%%%%%%%%%%%%%%%%%%%%%%%%%%%%%%%%%%%%%%%%%%%%%%%%%%%
% Results
%%%%%%%%%%%%%%%%%%%%%%%%%%%%%%%%%%%%%%%%%%%%%%%%%%%%%%%%%%%%%%%%%%%%%%%%%%%%%%%%%%%%%%%%%%%%%%%%%%%%%%%%%%%%%%%%%%%%%%%%%%
%%%%%%%%%%%%%%%%%%%%%%%%%%%%%%%%%%%%%%%%%%%%%%%%%%%%%%%%%%%%%%%%%%%%%%%%%%%%%%%%%%%%%%%%%%%%%%%%%%%%%%%%%%%%%%%%%%%%%%%%%%
\section{Results}\label{sec:results}
%%%%%%%%%%%%%%%%%%%%%%%%%%%%%%%%%%%%%%%%%%%%%%%%%%%%%%%%%%%%%%%%%%%%%%%%%%%%%%%%%%%%%%%%%%%%%%%%%%%%%%%%%%%%%%%%%%%%%%%%%%
%%%%%%%%%%%%%%%%%%%%%%%%%%%%%%%%%%%%%%%%%%%%%%%%%%%%%%%%%%%%%%%%%%%%%%%%%%%%%%%%%%%%%%%%%%%%%%%%%%%%%%%%%%%%%%%%%%%%%%%%%%
%%%%%%%%%%%%%%%%%%%%%%%%%%%%%%%%%%%%%%%%%%%%%%%%%%%%%%%%%%%%%%%%%%%%%%%%%%%%%%%%%%%%%%%%%%%%%%%%%%%%%%%%%%%%%%%%%%%%%%%%%%
%%%%%%%%%%%%%%%%%%%%%%%%%%%%%%%%%%%%%%%%%%%%%%%%%%%%%%%%%%%%%%%%%%%%%%%%%%%%%%%%%%%%%%%%%%%%%%%%%%%%%%%%%%%%%%%%%%%%%%%%%%
\subsection{Ground-state properties}
%%%%%%%%%%%%%%%%%%%%%%%%%%%%%%%%%%%%%%%%%%%%%%%%%%%%%%%%%%%%%%%%%%%%%%%%%%%%%%%%%%%%%%%%%%%%%%%%%%%%%%%%%%%%%%%%%%%%%%%%%%
%%%%%%%%%%%%%%%%%%%%%%%%%%%%%%%%%%%%%%%%%%%%%%%%%%%%%%%%%%%%%%%%%%%%%%%%%%%%%%%%%%%%%%%%%%%%%%%%%%%%%%%%%%%%%%%%%%%%%%%%%%
Since the number of electrons in our systems is large, we have to find some compromise and correlate the chemically most important electrons.  
Accordingly, we examined a number of correlated electrons distributed over different subshells in the ThO and ThS molecules as shown in Table~\ref{tbl:correlated-shells}. 
One can see that increasing the number of correlated electrons in both molecules leads to rather different atomic contributions. 
In the valence part of ThO only the 2$s$ and 2$p$ atomic oxygen spinors are present, while in ThS the contribution of sulfur covers up to four (2$s$, 2$p$, 3$s$, and 3$p$) atomic sulfur spinors in the calculations with 58 electrons. 
%%%%%%%%%%%%%%%%%%%%%%%%%%%%%%%%%%%%%%%%%%%%%%%%%%%%%%%%%%%%%%%%%%%%%%%%%%%%%%%%%%%%%%%%%%%%%%%%%%%%%%%%%%%%%%%%%%%%%%%%%%
%  Table 0
%%%%%%%%%%%%%%%%%%%%%%%%%%%%%%%%%%%%%%%%%%%%%%%%%%%%%%%%%%%%%%%%%%%%%%%%%%%%%%%%%%%%%%%%%%%%%%%%%%%%%%%%%%%%%%%%%%%%%%%%%%
\begin{table}[t!]\caption{Composition of correlated spinors for various number of electrons in ThO and ThS. N$_{\text{corr}}$ denotes the number of correlated electrons.}
\label{tbl:correlated-shells}
\begin{center}
\begin{tabular}{ *2{c}}  
    \hline
    \hline
            N$_{\text{corr}}$ &Atomic contributions in occupied spinors\\ \hline
            & \\ 
            ThO & \\ \hline 
            & \\ 
            28    & Th: 5$d$, 6$s$, 6$p$, 7$s$ \\
                  & O: 2$s, $2$p$              \\ 
            34    & Th: 5$p$, 5$d$, 6$s$, 6$p$, 7$s$ \\
                  & O: 2$s$, 2$p$              \\ 
            50    & Th: 4$f$, 5$s$, 5$p$, 5$d$, 6$s$, 6$p$, 7$s$ \\
                  & O: 2$s$, 2$p$              \\ 
            ThS & \\\hline 
            &\\ 
             28  & Th: 5$d$, 6$s$, 6$p$, 7$s$\\
                 & S: 3$s, $3$p$\\ 
             34  & Th: 5$d$, 6$s$, 6$p$, 7$s$\\
                 & S: 2$p$, 3$s$, 3$p$ \\
             44  & Th: 5$p$, 5$d$, 6$s$, 6$p$, 7$s$\\
                 & S: 2$s$, 2$p$, 3$s$, 3$p$ \\
             58  & Th: 4$f$, 5$p$, 5$d$, 6$s$, 6$p$, 7$s$\\
                 & S: 2$s$, 2$p$, 3$s$, 3$p$ \\
               \hline
    \hline
\end{tabular}
\end{center}
\end{table}
%%%%%%%%%%%%%%%%%%%%%%%%%%%%%%%%%%%%%%%%%%%%%%%%%%%%%%%%%%%%%%%%%%%%%%%%%%%%%%%%%%%%%%%%%%%%%%%%%%%%%%%%%%%%%%%%%%%%%%%%%%
% End of Table 1
%%%%%%%%%%%%%%%%%%%%%%%%%%%%%%%%%%%%%%%%%%%%%%%%%%%%%%%%%%%%%%%%%%%%%%%%%%%%%%%%%%%%%%%%%%%%%%%%%%%%%%%%%%%%%%%%%%%%%%%%%%

%%%%%%%%%%%%%%%%%%%%%%%%%%%%%%%%%%%%%%%%%%%%%%%%%%%%%%%%%%%%%%%%%%%%%%%%%%%%%%%%%%%%%%%%%%%%%%%%%%%%%%%%%%%%%%%%%%%%%%%%%%
%  Table 2
%%%%%%%%%%%%%%%%%%%%%%%%%%%%%%%%%%%%%%%%%%%%%%%%%%%%%%%%%%%%%%%%%%%%%%%%%%%%%%%%%%%%%%%%%%%%%%%%%%%%%%%%%%%%%%%%%%%%%%%%%%
\begin{table}[t!]\caption{The influence of correlated number of electrons (N$_{\text{corr}}$) on the ground state bond lengths ($\mathrm {r_e}$), vibrational frequencies ($\mathrm {\omega_e}$), and force constants ($\mathrm{f}$) of ThO and ThS using the dyall.v3z basis set for all elements.}
\label{tbl:numer-correl-el}
\begin{center}
\begin{tabular}{ *1{l}*5{c}}  
    \hline
    \hline
            Method& N$_{\text{corr}}$& $\mathrm {r_e}$ [\AA] & $\mathrm {\omega_e}$ [cm$^{-1}$]& $\mathrm{f}$ [N/m]\\ \hline
            & &   & &\\ 
            ThO & & & &\\\hline 
            & &   & &&\\ 
            CCSD   & 28&1.838 & 921&748.6\\
            CCSD   & 34& 1.837 & 922&749.6\\
            CCSD   & 50& 1.837 & 922&749.3\\
            & &   & &&\\ 
            CCSD(T)& 28& 1.848 & 894&704.0\\
            CCSD(T)& 34& 1.848 & 894&705.3\\
            CCSD(T)& 50& 1.848 & 894&705.1\\
            & &   & &&\\ 
            ThS & &  & &&\\\hline 
            & &   & &&\\ 
            CCSD   & 28&  2.348 &496&407.3 \\
            CCSD   & 34&  2.348 &496&407.1\\
            CCSD   & 44&  2.347 &496&407.9\\
            CCSD   & 58&  2.347 &496&408.0\\
            & &   & &&\\ 
            CCSD(T)& 28&  2.357 &482&384.4 \\
            CCSD(T)& 34&  2.356 &482&384.0\\
            CCSD(T)& 44&  2.355 &482&385.1\\
            CCSD(T)& 58&  2.355 &483&385.1\\
    \hline
    \hline
\end{tabular}
\end{center}
\end{table}
%%%%%%%%%%%%%%%%%%%%%%%%%%%%%%%%%%%%%%%%%%%%%%%%%%%%%%%%%%%%%%%%%%%%%%%%%%%%%%%%%%%%%%%%%%%%%%%%%%%%%%%%%%%%%%%%%%%%%%%%%%
% End of Table 2
%%%%%%%%%%%%%%%%%%%%%%%%%%%%%%%%%%%%%%%%%%%%%%%%%%%%%%%%%%%%%%%%%%%%%%%%%%%%%%%%%%%%%%%%%%%%%%%%%%%%%%%%%%%%%%%%%%%%%%%%%%
ThO and ThS ground state spectroscopic parameters calculated from the CCSD and CCSD(T) methods and various number of correlated electrons are reported in Table~\ref{tbl:numer-correl-el}. 
It is evident from Table~\ref{tbl:numer-correl-el} that for ThO it is sufficient to correlate only 34 electrons (5$p$, 5$d$, 6$s$, 6$p$, and 7$s$ spinors of thorium as well as  2$s$ and 2$p$ spinors of oxygen) as a larger number of correlated electrons does not effect spectroscopic constants. 
The only change in spectroscopic constants with respect to calculations performed with 50 electrons can be seen in the force constant. 
However, the difference is almost negligible. 
The need for correlating similar amount of electrons in the electronic structure of ThO has been recently discussed by Skripnikov and Titov~\cite{Skripnikov-ThO-2015,Skripnikov-ThO-2016}. 
  
For the ThS molecule, the optimal number of correlated electrons seems to be 34 as correlating additional electrons does not effect spectroscopic parameters considerably.  
The spinors of these calculations have similar atomic composition as the set of ThO with the same number of electrons, with the only difference that the thorium 5$p$ spinors are substituted by the presence of sulfur 2$p$ atomic spinors (cf. Table~\ref{tbl:correlated-shells} for comparison).
It is worth notifying that the presence of thorium 4$f$ atomic spinors in correlated calculations does not affect the spectroscopic constants of ThO and ThS. 
Hence, in all other calculations presented in this paper we correlated 34 electrons (32 for doubly charged molecules, \textit{vide infra}) in the coupled cluster approach.   
Further analysis of the ThO Dirac--Hartree--Fock molecular spinors shows that the largest mixing occurs for the thorium 6$p$ as well as the oxygen 2$s$ and 2$p$ atomic spinors. 
In the ThS molecule, the largest mixing is observed for the thorium 6$d$ (and to a lesser extend 6$p$), and the sulfur 3$s$ and 3$p$ atomic spinors.  
Therefore, inspection of the ThO and ThS Dirac--Hartree--Fock spinors confirms the earlier theoretical studies suggesting a triple bond.  
%%%%%%%%%%%%%%%%%%%%%%%%%%%%%%%%%%%%%%%%%%%%%%%%%%%%%%%%%%%%%%%%%%%%%%%%%%%%%%%%%%%%%%%%%%%%%%%%%%%%%%%%%%%%%%%%%%%%%%%%%%
%  Table 3
%%%%%%%%%%%%%%%%%%%%%%%%%%%%%%%%%%%%%%%%%%%%%%%%%%%%%%%%%%%%%%%%%%%%%%%%%%%%%%%%%%%%%%%%%%%%%%%%%%%%%%%%%%%%%%%%%%%%%%%%%%
\begin{table}[t!]\caption{The influence of the basis set size on the ground state bond lengths ($\mathrm {r_e}$), vibrational frequencies ($\mathrm {\omega_e}$), and force constants ($\mathrm{f}$) of ThO and ThS correlating 34 electrons. The CBS limit was obtained through the extrapolation of total Dirac--Hartee--Fock energies with 2,3, and 4 cardinal numbers and of correlation energies with 3 and 4 cardinal numbers. }
\label{tbl:basis-extrapol}
\begin{center}
\begin{tabular}{ *1{l}*5{c}}  
    \hline
    \hline
            Method& Basis Th/$\{$O,S$\}$&  $\mathrm {r_e}$ [\AA] & $\mathrm {\omega_e}$ [cm$^{-1}$]& $\mathrm{f}$ [N/m]\\ \hline
            & &   & &\\ 
            ThO & & & &\\\hline 
            & &   & &\\ 
            CCSD        &dyall.v3z/dyall.v3z&1.837 &922 &749.6\\
            CCSD        & dyall.v3z/cc-pVTZ &1.837 &924 &752.2\\
            CCSD        & dyall.v4z/cc-pVQZ &1.833 &924 &752.7\\
            CCSD        & CBS               &1.831 &924 &753.3\\

            CCSD(T)     &dyall.v3z/dyall.v3z&1.848 &894 &705.3\\ 
            CCSD(T)     & dyall.v3z/cc-pVTZ &1.848 &896 &708.4\\
            CCSD(T)     & dyall.v4z/cc-pVQZ &1.844 &896 &708.0 \\
            CCSD(T)     &               CBS &1.842 &896 &707.8\\
            & &   & &&\\ 
            ThS & &  & &&\\\hline 
            & &   & &&\\
            CCSD        &dyall.v3z/dyall.v3z&2.348 &496 &407.1\\ 
            CCSD        & dyall.v3z/cc-pVTZ &2.348 &497 &409.3\\
            CCSD        & dyall.v4z/cc-pVQZ &2.341 &499 &412.8\\
            CCSD        &               CBS &2.337 &501 &415.9\\
           
            CCSD(T)     &dyall.v3z/dyall.v3z&2.356 &482 &384.0\\
            CCSD(T)     & dyall.v3z/cc-pVTZ &2.356 &483 &386.5\\
            CCSD(T)     & dyall.v4z/cc-pVQZ &2.349 &486 &390.4\\
            CCSD(T)     &                CBS&2.345 &488 &393.6\\
    \hline
    \hline
\end{tabular}
\end{center}
\end{table}
%%%%%%%%%%%%%%%%%%%%%%%%%%%%%%%%%%%%%%%%%%%%%%%%%%%%%%%%%%%%%%%%%%%%%%%%%%%%%%%%%%%%%%%%%%%%%%%%%%%%%%%%%%%%%%%%%%%%%%%%%%
% End of Table 3
%%%%%%%%%%%%%%%%%%%%%%%%%%%%%%%%%%%%%%%%%%%%%%%%%%%%%%%%%%%%%%%%%%%%%%%%%%%%%%%%%%%%%%%%%%%%%%%%%%%%%%%%%%%%%%%%%%%%%%%%%%

In Table~\ref{tbl:basis-extrapol}, we analyzed the influence of basis set size on the ground state spectroscopic constants of ThO and ThS. 
It is evident from these results that the triple zeta quality basis set (despite of its type) provides already very good results for both molecules.  
Comparing triple zeta results with those extrapolated to the basis set limit shows only minor differences in spectroscopic constants (see Table~\ref{tbl:basis-extrapol}). 
Exception is the ThS equilibrium bond length, which changes up to 0.011~\AA~for both CCSD and CCSD(T). 
At the same time, the differences between vibrational frequencies are overall small and do not exceed few reciprocal centimeters.  

In Table~\ref{tbl:speconstho} we compare the new ground state theoretical results for ThO to the existing experimental and theoretical data available in the literature.  
%%%%%%%%%%%%%%%%%%%%%%%%%%%%%%%%%%%%%%%%%%%%%%%%%%%%%%%%%%%%%%%%%%%%%%%%%%%%%%%%%%%%%%%%%%%%%%%%%%%%%%%%%%%%%%%%%%%%%%%%%%
%  Table 4
%%%%%%%%%%%%%%%%%%%%%%%%%%%%%%%%%%%%%%%%%%%%%%%%%%%%%%%%%%%%%%%%%%%%%%%%%%%%%%%%%%%%%%%%%%%%%%%%%%%%%%%%%%%%%%%%%%%%%%%%%%
\begin{table}[t!]
\begin{center}
\small
\caption{Equilibrium bond lengths ($\mathrm {r_e}$) and vibrational frequencies ($\mathrm {\omega_e}$) of the $^1\Sigma^+$ ground-state ThO.}
\label{tbl:speconstho}
\begin{tabular}{ ccc}  
     
    \hline
    \hline
    Method & $\mathrm{r_e}$ [\AA] & $\mathrm{\omega_e} [\mathrm{cm}^{-1}]$ \\ \hline
    \textbf{Experimental}  & & \\
     PFI-ZEKE (gas phase)~\cite{exp_goncharov1} & 1.840 &  \\
     MW (gas phase)~\cite{exp_dewberry} & 1.840 &  \\
     Electron Spec. (gas phase)~ \cite{exp_edvinsson} & & 896  \\
     IR Ne matrix~\cite{exp_andrews_jacs} &  &887 \\
     IR Ar matrix~\cite{exp_kushto} & & 879 \\ \hline
    \textbf{Theoretical} & &  \\ 
    ECP CASSCF~\cite{Marian_tho_1988} & 1.927 & 847\\
    DFT/B3PW91~\cite{exp_goncharov1} & 1.846 & 898 \\
    CASSCF/MRCI~ \cite{dolg_pseudo}& 1.862 & 867 \\
    Spin Orbit CASPT2~\cite{theo_dk3_paulovic} & 1.862 & 856 \\ 
    Spin Free CASPT2~\cite{kovacs_freqs} & 1.861 & 879\\ \hline
    \textbf{Present work}  & & \\
    CCSD/dyall.v3z & 1.837 & 922  \\
    CCSD/CBS & 1.831 & 924  \\
    CCSD(T)/dyall.v3z & 1.848 & 894 \\
    CCSD(T)/CBS & 1.842 & 896 \\
    \hline
    \end{tabular}   
 \end{center} 
 \end{table}
%%%%%%%%%%%%%%%%%%%%%%%%%%%%%%%%%%%%%%%%%%%%%%%%%%%%%%%%%%%%%%%%%%%%%%%%%%%%%%%%%%%%%%%%%%%%%%%%%%%%%%%%%%%%%%%%%%%%%%%%%%
%  End of Table 4
%%%%%%%%%%%%%%%%%%%%%%%%%%%%%%%%%%%%%%%%%%%%%%%%%%%%%%%%%%%%%%%%%%%%%%%%%%%%%%%%%%%%%%%%%%%%%%%%%%%%%%%%%%%%%%%%%%%%%%%%%%
One can see that our CCSD and CCSD(T) results are very close to experiment, outperforming the standard CASSCF/MRCI and CASPT2 approaches for bond lengths and vibrational frequencies. 
Extrapolation to the basis set limit brings the CCSD(T) bond length (1.842~\AA) closer to experimental value of 1.840~\AA. 
Both CCSD vibrational frequencies overestimate the experimental value by approximately 30 cm$^{-1}$. 
The triples correction on top of CCSD brings the characteristic vibrations very close to experimentally determined values (879--896  cm$^{-1}$). 
Our findings are in line with recent work of Skripnikov and Titov~\cite{Skripnikov-ThO-2015} who estimated contributions form triple and higher rank excitations in the ThO  molecule to 5\%. 

Table~\ref{tbl:speconsths} collects the ground state bond lengths and vibrational frequencies for ThS.  
%%%%%%%%%%%%%%%%%%%%%%%%%%%%%%%%%%%%%%%%%%%%%%%%%%%%%%%%%%%%%%%%%%%%%%%%%%%%%%%%%%%%%%%%%%%%%%%%%%%%%%%%%%%%%%%%%%%%%%%%%%
%  Table 5
%%%%%%%%%%%%%%%%%%%%%%%%%%%%%%%%%%%%%%%%%%%%%%%%%%%%%%%%%%%%%%%%%%%%%%%%%%%%%%%%%%%%%%%%%%%%%%%%%%%%%%%%%%%%%%%%%%%%%%%%%%
\begin{table}[t!]
\begin{center}
\small
\caption{Equilibrium bond lengths ($\mathrm {r_e}$) and vibrational frequencies ($\mathrm {\omega_e}$) of the $^1\Sigma^+$ ground-state ThS.}
 \label{tbl:speconsths}
\begin{tabular}{ *3{c}}  
     
    \hline
    \hline
     Method & $\mathrm {r_e} [$\AA$]$ & $\mathrm{\omega_e} [\mathrm{cm}^{-1}]$ \\ \hline
    \textbf{Experimental}  & & \\
    IR~\cite{ThS_liang_andrews_2002} &  & 475 \\
    Electron Spec. (gas phase)~\cite{ThS_Barlett_Antonov_Heaven2013} & & 479(1)  \\
    \textbf{Theoretical}  & &  \\
    B3LYP~\cite{Pereira-ths} & 2.349 & 481  \\
    B3PW91~\cite{ThS_liang_andrews_2002} & 2.341 & 479  \\
    CASSCF/MRCI~\cite{ThS_Barlett_Antonov_Heaven2013} & 2.363 & 477  \\
    \textbf{Present work} &  & \\
    CCSD/dyall.v3z &2.348  &496   \\
    CCSD/CBS & 2.337 & 501  \\
    CCSD(T)/dyall.v3z & 2.356  &482   \\
    CCSD(T)/CBS &2.345  &488   \\
    \hline
    \end{tabular}   
 \end{center} 
 \end{table}
%%%%%%%%%%%%%%%%%%%%%%%%%%%%%%%%%%%%%%%%%%%%%%%%%%%%%%%%%%%%%%%%%%%%%%%%%%%%%%%%%%%%%%%%%%%%%%%%%%%%%%%%%%%%%%%%%%%%%%%%%
%  End of Table 5
%%%%%%%%%%%%%%%%%%%%%%%%%%%%%%%%%%%%%%%%%%%%%%%%%%%%%%%%%%%%%%%%%%%%%%%%%%%%%%%%%%%%%%%%%%%%%%%%%%%%%%%%%%%%%%%%%%%%%%%%%%
Our CCSD and CCSD(T) vibrational frequencies are in very good agreement with the experimental values, with CCSD(T) being somehow closer.
Specifically, the CCSD(T) values of 483 and 488 cm$^{-1}$ fit quite well the experimental values of 475 and 479 cm$^{-1}$. 
The difference between the CCSD and CCSD(T) vibrational frequencies is similar as observed for ThO and amounts to 30 cm$^{-1}$. 
In both molecules the CCSD(T) bond lengths are approximately 0.01~\AA~longer than those obtained from the CCSD calculations. 

Comparing the ground-state spectroscopic constants of ThO and ThS, we can see that the Th--S bond length is ca. 0.51~\AA~longer than the corresponding Th--O bond. 
Such a considerable difference can be explained by the approximately two times larger atomic sphere of sulfur.  
The force constants of ThO are roughly twice the value of ThS (705 vs. 385 \textrm{N/m}), thus the strength of the Th--O bond is approximately twice the strength of Th--S.  
For both molecules, the CCSD(T) spectroscopic constants (compare Tables~\ref{tbl:speconstho} and~\ref{tbl:speconsths}) agrees better with experimental and 
other theoretical data. 
Moderately large values of the CCSD T$_1$-diagnostics, 0.030 and 0.034 for ThO and ThS, respectively, reveal potential problems for perturbative corrections on top of CCSD, such as CCSD(T). However, the good agreement of ThO and ThS vibrational frequencies with experimental values suggests adequacy of the CCSD(T) model at least around the equilibrium structures. 
%%%%%%%%%%%%%%%%%%%%%%%%%%%%%%%%%%%%%%%%%%%%%%%%%%%%%%%%%%%%%%%%%%%%%%%%%%%%%%%%%%%%%%%%%%%%%%%%%%%%%%%%%%%%%%%%%%%%%%%%%%
%%%%%%%%%%%%%%%%%%%%%%%%%%%%%%%%%%%%%%%%%%%%%%%%%%%%%%%%%%%%%%%%%%%%%%%%%%%%%%%%%%%%%%%%%%%%%%%%%%%%%%%%%%%%%%%%%%%%%%%%%%
\subsection{Excited-state properties}
%%%%%%%%%%%%%%%%%%%%%%%%%%%%%%%%%%%%%%%%%%%%%%%%%%%%%%%%%%%%%%%%%%%%%%%%%%%%%%%%%%%%%%%%%%%%%%%%%%%%%%%%%%%%%%%%%%%%%%%%%%
%%%%%%%%%%%%%%%%%%%%%%%%%%%%%%%%%%%%%%%%%%%%%%%%%%%%%%%%%%%%%%%%%%%%%%%%%%%%%%%%%%%%%%%%%%%%%%%%%%%%%%%%%%%%%%%%%%%%%%%%%%
%%%%%%%%%%%%%%%%%%%%%%%%%%%%%%%%%%%%%%%%%%%%%%%%%%%%%%%%%%%%%%%%%%%%%%%%%%%%%%%%%%%%%%%%%%%%%%%%%%%%%%%%%%%%%%%%%%%%%%%%%%
%%%%%%%%%%%%%%%%%%%%%%%%%%%%%%%%%%%%%%%%%%%%%%%%%%%%%%%%%%%%%%%%%%%%%%%%%%%%%%%%%%%%%%%%%%%%%%%%%%%%%%%%%%%%%%%%%%%%%%%%%%
\subsubsection{ThO}
%%%%%%%%%%%%%%%%%%%%%%%%%%%%%%%%%%%%%%%%%%%%%%%%%%%%%%%%%%%%%%%%%%%%%%%%%%%%%%%%%%%%%%%%%%%%%%%%%%%%%%%%%%%%%%%%%%%%%%%%%%
%%%%%%%%%%%%%%%%%%%%%%%%%%%%%%%%%%%%%%%%%%%%%%%%%%%%%%%%%%%%%%%%%%%%%%%%%%%%%%%%%%%%%%%%%%%%%%%%%%%%%%%%%%%%%%%%%%%%%%%%%%

Table~\ref{tbl:adiabatic-excit-tho} lists the low-lying part of adiabatic spectrum of the ThO molecule. 
Our FSCC spectroscopic parameters are compared to the SO-CASPT2 numbers and when possible also to experiment. 
%%%%%%%%%%%%%%%%%%%%%%%%%%%%%%%%%%%%%%%%%%%%%%%%%%%%%%%%%%%%%%%%%%%%%%%%%%%%%%%%%%%%%%%%%%%%%%%%%%%%%%%%%%%%%%%%%%%%%%%%%%
%  Table 6
%%%%%%%%%%%%%%%%%%%%%%%%%%%%%%%%%%%%%%%%%%%%%%%%%%%%%%%%%%%%%%%%%%%%%%%%%%%%%%%%%%%%%%%%%%%%%%%%%%%%%%%%%%%%%%%%%%%%%%%%%%
\begin{table*}[ht!]
\begin{center}
\caption{Spin--orbit electronic spectrum of the ThO molecule. Adiabatic excitation energies, $\mathrm{T_e}$ (in cm$^{-1}$), bond lengths, $\mathrm{r_e}$ (in~\AA), and vibrational frequencies, $\mathrm{\omega_e}$ (in cm$^{-1}$).}\label{tbl:adiabatic-excit-tho}
\small
\begin{tabular}{l |ccc| ccc| ccc| ccc}
\hline
\hline
\multicolumn{1}{c|}{State}&
\multicolumn{3}{c|}{IHFSCC(1,1)}&
\multicolumn{3}{c|}{IHFSCC(0,2)}&
\multicolumn{3}{c|}{SO-CASPT2~\cite{theo_dk3_paulovic}}&
\multicolumn{3}{c}{Experiment}\\
\hline
 $\Omega$ & $\mathrm{T_e}$[cm$^{-1}$] &$\mathrm{r_e}$[\AA] &$\mathrm{\omega_e}$[cm$^{-1}$] & $\mathrm{T_e}$[cm$^{-1}$] &$\mathrm{r_e}$[\AA] &$\mathrm{\omega_e}$[cm$^{-1}$] 
 & $\mathrm{T_e}$[cm$^{-1}$] &$\mathrm{r_e}$[\AA] &$\mathrm{\omega_e}$[cm$^{-1}$] & $\mathrm{T_e}$[cm$^{-1}$] &$\mathrm{r_e}$[\AA] &$\mathrm{\omega_e}$[cm$^{-1}$] \\

\hline
 0$^+$(X)  &     0&1.837&922&0     &1.841 & 922&0     &1.866&856 &0     &1.840&896 \\
 1 (H)     & 5 168&1.854&885& 6 017&1.855 & 885& 5 549&1.882&823 &5 317 &1.858&857 \\
 2 (Q)     & 6 086&1.853&886& 6 866&1.854 & 886& 6 693&1.880&828 &6 128 &1.856&858 \\
 3 (W)     & 7 694&1.852&887& 8 438&1.852 & 889& 8 408&1.878&835 &8 600&-&- \\
 0$^-$     &10 701&1.861&857&10 911&1.857 & 882&10 370&1.901&784 &-&-&- \\
 0$^+$(A)  &11 699&1.862&910&11 292&1.857 & 882&10 388&1.902&783 &10 601&1.867&846 \\
 1 (B)     &      &     &   &12 056&1.859 & 879&11 181&1.905&776 &11 129&1.864&843\\
 2         &12 803&1.849&885&12 732&1.852 & 886&12 891&1.900&774 &-&-&- \\
 1 (C)     &14 451&1.866&859&16 188&1.864 & 869&14 112&1.914&779 &14 490&1.870&825\\
 2         &14 997&1.859&872&14 533&1.857 & 883&14 640&1.872&853 &-&-&- \\
 1 (D)     &-     &    -&  -&17 644&1.862 & 874&19 813&1.914&701 &15 946&1.866&839\\
 0$^-$     &16 982&1.888&822&18 016&1.868 & 855&20 188&1.879&866 &-&-&- \\
 0$^+$(E)  &14 370&1.868&855&17 280&1.859 & 875&17 912&1.902&781 &16 320&1.867&829 \\
 2 (G)     &-     &    -&  -&     -&     -&   -&17 339&1.920&759 &18 010&1.882&809\\

\hline
\end{tabular}
\end{center}
\end{table*}
%%%%%%%%%%%%%%%%%%%%%%%%%%%%%%%%%%%%%%%%%%%%%%%%%%%%%%%%%%%%%%%%%%%%%%%%%%%%%%%%%%%%%%%%%%%%%%%%%%%%%%%%%%%%%%%%%%%%%%%%%%
%  End of Table 6
%%%%%%%%%%%%%%%%%%%%%%%%%%%%%%%%%%%%%%%%%%%%%%%%%%%%%%%%%%%%%%%%%%%%%%%%%%%%%%%%%%%%%%%%%%%%%%%%%%%%%%%%%%%%%%%%%%%%%%%%%%
The calculated excitation energies of ThO are mainly dominated by electron transfer from the occupied 7$s$ spinor to the unoccupied 6$d$ and 7$p$ spinors of thorium. 
The composition of virtual spinors changes, however, with the bond distance. 
In general, the electronic spectrum of ThO can be divided into three blocks. 

The first block of the electronic spectrum covers excitations to the 6$d$ spinors in the range of 5 000--8 500 cm$^{-1}$. 
In the perturbative spin--orbit CASPT2 calculations~\cite{theo_dk3_paulovic} these excitations have mainly $^3\Delta$ character. 
Our Mulliken-based population analysis of virtual spinors confirms the leading contribution form the $\delta$-type spinors in this part of the spectrum. 
It is worth noting that in the relativistic Dirac equation used here, spin is not a good quantum number and therefore individual contributions from singlet and triplet cannot be anticipated. 
In this part of the spectrum the bond lengths are elongated by about 0.01--0.02~\AA~ and vibrational frequencies lowered by approximately 40 cm$^{-1}$ with respect to the ground state reference. This is true for all theoretical and experimental spectra (see Table~\ref{tbl:adiabatic-excit-tho} for more details). 

The second block ranges from 10 000 to 13 000 cm$^{-1}$ and includes electron transfer to 7$p$ spinors. 
In the perturbative spin--orbit CASPT2 calculations~\cite{theo_dk3_paulovic} these excited states are characterized by leading contributions from the $^3\Pi$ state. 
All equilibrium bond lengths for these excited states are slightly longer than in the first block of the ThO spectrum (see Table~\ref{tbl:adiabatic-excit-tho}). 
The intrinsic part of this spectrum are quasi-degenerate $0^{-}$ and $0^{+}$ excited states. 
The IHFSCC(1,1) $0^{+}$ excited state potential energy surface deviates from quadratic shape and exhibits a double minima.  
As a result the vibrational frequency of this particular state is larger than its $0^{-}$ counterpart (910 vs. 857 cm$^-1)$. 
Such a double shoulder feature of the $0^{+}$ excited state is not present in the IHFSCC(0,2) approach. 
In that case, both $0^{-}$ and $0^{+}$ excited states are characterized by the same vibrational frequency and almost identical quadratic shape of potential energy curves. 
 
The third block contains all the remaining excitations up to 18 000 cm$^{-1}$ and covers electronic transitions to the mixed 6$d$ and 7$p$ molecular spinors. 
In this part of the spectrum we see a lot of distortions in the bond lengths and the values of vibrational frequencies depending on the theoretical method used. 
In general, the IHFSCC(1,1) excitation energies are too low and the IHFSCC(0,2) excitation energies too high compared with experimental values.  
The 2 (G) excited state present in experiment and CASPT2 (bearing significant $\Phi$ character) is not calculated in our FSCC spectrum as it was impossible to include the virtual spinors from the 5$f$ shell in the main model space.  
 
Comparing our IHFSCC(1,1) and IHFSCC(0,2) spectroscopic constants we see that both methods give similar results. 
The agreement between these two approaches is smallest in the lower part of the ThO spectrum and increases towards higher lying excited states. 
It is worth noticing that in the low-lying part of the electronic spectrum the compositions of the IHFSCC(1,1) and IHFSCC(0,2) model spaces (in terms of wave function character) are essentially identical to each other as shown in Table S5 of the ESI. 
The leading wave function character of the model spaces changes, however, for higher-lying excitation (see Table S5 for more details) as a result of different composition of higher-lying virtual spinors in the ThO and ThO$^{2+}$ reference states.  
Nevertheless, both methods predict the same order of excites states, while the differences in excitation energies usually do not exceed 1 000 cm$^{-1}$. 
The largest discrepancy is observed for the 1 (C) and 0$^{+}$ (E) excited states and amount to approximately 2 000 cm$^{-1}$. 
Finally, we should note that the 1 (B) and 1 (D) excited states are not reported for the IHFSCC(1,1) spectrum as they have significant contributions from the (undesired) $P_i$ space. 
Such excitation energies might not have physical meaning and should not be trusted.  

All FSCC spectroscopic parameters (bond lengths, excitation energies, and vibrational frequencies) agree rather well with experimental data listed in Table~\ref{tbl:adiabatic-excit-tho}. 
Specifically, the differences  do not exceed 50 cm$^{-1}$ in vibrational frequencies and 0.005~\AA~ in bond lengths. 
Based on our analysis of the ground state spectroscopic constants one might expect that to a large extent these discrepancies are caused by the lack of triple excitations in our model (\textit{vide supra} discussion on the CCSD and CCSD(T) results).   
In general, the IHFSCC results predict shorter bond lengths and higher vibrational frequencies than those obtained from SO-CASPT2. 
Such trends in SO-CASPT2 spectra have already been observed for other actinide species~\cite{pawel2,Cs2UO2Cl4-Gomes}. 
%%%%%%%%%%%%%%%%%%%%%%%%%%%%%%%%%%%%%%%%%%%%%%%%%%%%%%%%%%%%%%%%%%%%%%%%%%%%%%%%%%%%%%%%%%%%%%%%%%%%%%%%%%%%%%%%%%%%%%%%%%
%%%%%%%%%%%%%%%%%%%%%%%%%%%%%%%%%%%%%%%%%%%%%%%%%%%%%%%%%%%%%%%%%%%%%%%%%%%%%%%%%%%%%%%%%%%%%%%%%%%%%%%%%%%%%%%%%%%%%%%%%%
\subsubsection{ThS}
%%%%%%%%%%%%%%%%%%%%%%%%%%%%%%%%%%%%%%%%%%%%%%%%%%%%%%%%%%%%%%%%%%%%%%%%%%%%%%%%%%%%%%%%%%%%%%%%%%%%%%%%%%%%%%%%%%%%%%%%%%
%%%%%%%%%%%%%%%%%%%%%%%%%%%%%%%%%%%%%%%%%%%%%%%%%%%%%%%%%%%%%%%%%%%%%%%%%%%%%%%%%%%%%%%%%%%%%%%%%%%%%%%%%%%%%%%%%%%%%%%%%%
The adiabatic excitation energies of the ThS molecule are listed in Table~\ref{tbl:adiabatic-excit-ths}. 
In general, the electronic spectrum of ThS bears a lot of similarities to the spectrum of ThO. 
Similar to ThO, also the ThS electronic spectrum can be divided into three characteristic blocks. 

The lowest part of the spectrum covers excitations from the 7$s$ atomic spinor to the 6$d$ spinors in the range of 2 500--8 000 cm$^{-1}$. 
Despite the fact that excitation energies have similar character (dominant $\delta$ contributions) and the same symmetry ($\Omega =$ 1, 2, and 3), they are lower in energy by approximately 2 000 cm$^{-1}$ than in the corresponding ThO molecule. 
Similar as in ThO, the optimal bond lengths for excited states are elongated by approximately 0.02~\AA~and vibrational frequencies lowered by about 20 cm$^{-1}$ with respect to the ground state reference. 

In the second part of ThS spectrum covering excitations within 8 500 and 12 000 cm$^{-1}$, the bond lengths are elongated by additional 0.02~\AA. 
In the ThS molecule, the separation between the first pair of 0$^+$ and 0$^-$ states is lowered down to 400 cm$^{-1}$ compared with 1 000 cm$^{-1}$ in the ThO molecule. 

The remaining part of the ThS excitation energies is included in the third part of the spectrum. 
As in ThO, also here we observe a stronger sensitivity of the applied IHFSCC variant on the bond lengths and vibrational frequencies of excited states.
The overall agreement between electronic spectra obtained from sector (1,1) and sector (0,2) of the Fock space is a little bit less satisfactory than for ThO. 
The largest difference between these two variants of the IHFSCC approach occurs in the lowest-lying part of the electronic spectrum and amounts to 2 000 cm$^{-1}$ (cf. Table~\ref{tbl:adiabatic-excit-ths}). 
This is very surprising as in this part of the electronic spectrum wave function compositions of the IHFSCC(1,1) and IHFSCC(0,2) model spaces are almost identical (see Table S6 of the ESI for more details).  
Thus, orbital relaxation effects seem to be more important for ThS$^{2+}$ than than for ThO$^{2+}$.  
One should also note that the SO-MRCI results lie somehow between the IHFSCC(0,2) and IHFSCC(1,1) excitation energies.  
%%%%%%%%%%%%%%%%%%%%%%%%%%%%%%%%%%%%%%%%%%%%%%%%%%%%%%%%%%%%%%%%%%%%%%%%%%%%%%%%%%%%%%%%%%%%%%%%%%%%%%%%%%%%%%%%%%%%%%%%%%
%  Table 7
%%%%%%%%%%%%%%%%%%%%%%%%%%%%%%%%%%%%%%%%%%%%%%%%%%%%%%%%%%%%%%%%%%%%%%%%%%%%%%%%%%%%%%%%%%%%%%%%%%%%%%%%%%%%%%%%%%%%%%%%%%
\begin{table*}[ht!]
\begin{center}
\caption{Spin--orbit electronic spectrum of the ThS molecule. Adiabatic excitation energies, $\mathrm{T_e}$ (in cm$^{-1}$), bond lengths, $\mathrm{r_e}$ (in~\AA), and vibrational frequencies, $\mathrm{\omega_e}$ (in cm$^{-1}$).}\label{tbl:adiabatic-excit-ths}
\small
\begin{tabular}{l |ccc| ccc| ccc}
\hline
\hline
\multicolumn{1}{c|}{State}&
\multicolumn{3}{c|}{IHFSCC(1,1)}&
\multicolumn{3}{c|}{IHFSCC(0,2)}&
\multicolumn{3}{c}{SO-MRCI~\cite{ThS_Barlett_Antonov_Heaven2013}}\\
\hline
 $\Omega$ &  $\mathrm{T_e}$[cm$^{-1}$] &$\mathrm{r_e}$[\AA] &$\mathrm{\omega_e}$[cm$^{-1}$] & $\mathrm{T_e}$[cm$^{-1}$] &$\mathrm{r_e}$[\AA] &$\mathrm{\omega_e}$[cm$^{-1}$] 
 & $\mathrm{T_e}$[cm$^{-1}$] &$\mathrm{r_e}$[\AA] &$\mathrm{\omega_e}$[cm$^{-1}$] \\

\hline
 0     &0     &2.348&497	&0 	&2.344&506&0     &2.363&477\\
 1     &2 616 &2.377&475	&4 624 	&2.367&492&3 940 &2.394&454\\
 2     &3 642 &2.376&475	&5 572 	&2.366&493&4 856 &2.393&453\\
 3     &5 344 &2.374&478	&7 197 	&2.364&495&5 811 &2.391&455\\
 0$^-$ &8 742 &2.400&452	&9 961	&2.374&483&-&-&-\\
 0$^+$ &9 108 &2.403&448	&10 291 &2.374&483&-&-&-\\
 1     &    - & -   &- 	&10 964	&2.377&478&-&-&-\\
 2     &11 145&2.390&437	&12 245	&2.369&486&-&-&-\\
 2     &12 723&2.395&484	&13 852	&2.375&481&-&-&-\\
 1     &-     &-    &- 	&14 696	&2.386&464&-&-&-\\
 0$^+$ &16 090&2.413&450	&15 655	&2.391&455&-&-&-\\
 0$^-$ &18 218&2.427&444	&16 675	&2.389&460&-&-&-\\

\hline
\end{tabular}
\end{center}
\end{table*}
%%%%%%%%%%%%%%%%%%%%%%%%%%%%%%%%%%%%%%%%%%%%%%%%%%%%%%%%%%%%%%%%%%%%%%%%%%%%%%%%%%%%%%%%%%%%%%%%%%%%%%%%%%%%%%%%%%%%%%%%%%
%  End of Table 7
%%%%%%%%%%%%%%%%%%%%%%%%%%%%%%%%%%%%%%%%%%%%%%%%%%%%%%%%%%%%%%%%%%%%%%%%%%%%%%%%%%%%%%%%%%%%%%%%%%%%%%%%%%%%%%%%%%%%%%%%%%
%%%%%%%%%%%%%%%%%%%%%%%%%%%%%%%%%%%%%%%%%%%%%%%%%%%%%%%%%%%%%%%%%%%%%%%%%%%%%%%%%%%%%%%%%%%%%%%%%%%%%%%%%%%%%%%%%%%%%%%%%%
%%%%%%%%%%%%%%%%%%%%%%%%%%%%%%%%%%%%%%%%%%%%%%%%%%%%%%%%%%%%%%%%%%%%%%%%%%%%%%%%%%%%%%%%%%%%%%%%%%%%%%%%%%%%%%%%%%%%%%%%%%
\subsection{Comparison to the isoelectronic ThF$^+$ molecule}
%%%%%%%%%%%%%%%%%%%%%%%%%%%%%%%%%%%%%%%%%%%%%%%%%%%%%%%%%%%%%%%%%%%%%%%%%%%%%%%%%%%%%%%%%%%%%%%%%%%%%%%%%%%%%%%%%%%%%%%%%%
%%%%%%%%%%%%%%%%%%%%%%%%%%%%%%%%%%%%%%%%%%%%%%%%%%%%%%%%%%%%%%%%%%%%%%%%%%%%%%%%%%%%%%%%%%%%%%%%%%%%%%%%%%%%%%%%%%%%%%%%%%
It is worth noting that the ThF$^+$ molecule, which is isolectronic with ThO and valence isoelectronic to ThS has also been investigated as a potential candidate for the electron EDM~\cite{Mayer-ThO-ThF,Baker-ThF}. 
In-depth theoretical studies of the ThF$^+$ electronic structure revealed its complex nature resulting from equi-energetic $\Omega=0^+$ ($^1\Sigma_0^+$) and $\Omega=1$ ($^3\Delta_1$) states~\cite{Baker-ThF,ThF+2015}. 
Specifically, the order of the two lowest lying states in ThF$^+$ is very sensitive to the applied electron correlation method, treatment of spin--orbit coupling, and the basis set size~\cite{ThF+2015}. 
In this respect the ThF$^+$ electronic structure is very different from ThO and ThS, in which the ground state is well separated from the $^3\Delta_1$ excited state. 
The ThF$^+$ molecule is also more sensitive to the basis set choice than ThO and ThS.   
%%%%%%%%%%%%%%%%%%%%%%%%%%%%%%%%%%%%%%%%%%%%%%%%%%%%%%%%%%%%%%%%%%%%%%%%%%%%%%%%%%%%%%%%%%%%%%%%%%%%%%%%%%%%%%%%%%%%%%%%%%
%%%%%%%%%%%%%%%%%%%%%%%%%%%%%%%%%%%%%%%%%%%%%%%%%%%%%%%%%%%%%%%%%%%%%%%%%%%%%%%%%%%%%%%%%%%%%%%%%%%%%%%%%%%%%%%%%%%%%%%%%%
\subsection{Ionization potentials and electron affinities}\label{sec:ipea}
%%%%%%%%%%%%%%%%%%%%%%%%%%%%%%%%%%%%%%%%%%%%%%%%%%%%%%%%%%%%%%%%%%%%%%%%%%%%%%%%%%%%%%%%%%%%%%%%%%%%%%%%%%%%%%%%%%%%%%%%%%
%%%%%%%%%%%%%%%%%%%%%%%%%%%%%%%%%%%%%%%%%%%%%%%%%%%%%%%%%%%%%%%%%%%%%%%%%%%%%%%%%%%%%%%%%%%%%%%%%%%%%%%%%%%%%%%%%%%%%%%%%%
Finally, we discuss the first ionization potential (IP$_1$) and electron affinity (EA$_1$) of the ThO and ThS species.  
The IP$_1$ and EA$_1$ were calculated from sectors (1,0) and (0,1) of the Fock space, respectively. 
Our IHFSCC values are compared to the existing theoretical and experimental data in Table~\ref{tbl:ip-ea}. 
The first IP of the ThO molecule is predicted to be 6.69 eV and agrees rather well with the newest and very accurate experimental value of 6.60 eV~\cite{exp_goncharov_ie} as well as with other relativistic calculations of Infante~\textit{et al.}~\cite{Infante_ies_tho} (cf. Table~\ref{tbl:ip-ea}). 
 The IP$_1$ of ThS is only 0.19 eV higher than in the the ThO molecule. Larger differences between ThO and ThS are observed for the EA$_1$ values and amount to 0.51 eV. 
%%%%%%%%%%%%%%%%%%%%%%%%%%%%%%%%%%%%%%%%%%%%%%%%%%%%%%%%%%%%%%%%%%%%%%%%%%%%%%%%%%%%%%%%%%%%%%%%%%%%%%%%%%%%%%%%%%%%%%%%%%
%  Table 8
%%%%%%%%%%%%%%%%%%%%%%%%%%%%%%%%%%%%%%%%%%%%%%%%%%%%%%%%%%%%%%%%%%%%%%%%%%%%%%%%%%%%%%%%%%%%%%%%%%%%%%%%%%%%%%%%%%%%%%%%%%
\begin{table}[h!]
\caption{The first ionization potential (IP$_1$) and electron affinity (EA$_1$) of the ThO and ThS molecules.}\label{tbl:ip-ea}
\begin{center}
     
\small
\begin{tabular}{cc | cc}
\hline
\hline
\multicolumn{2}{c|}{ThO}&
\multicolumn{2}{c}{ThS}\\
    \hline
    \hline

 Method & IP$_1$ (eV) & Method &IP$_1$ (eV)\\
    \hline

 IHFSCC(1,0) & 6.69  & IHFSCC(1,0) &6.88  \\
CCSD~\cite{Infante_ies_tho}    &6.41&B3LYP~\cite{Pereira-ths}&6.61 \\
CCSD(T)~\cite{Infante_ies_tho} &6.55&MPW1PW91~\cite{Pereira-ths}&6.46 \\
SO-CASPT2~\cite{Infante_ies_tho}&6.56&& \\
Experiment~\cite{exp_goncharov_ie}&6.60&& \\
&&& \\\hline
 Method & EA$_1$ (eV) & Method &EA$_1$ (eV)\\
    \hline
 IHFSCC(0,1) & 0.58  & IHFSCC(0,1) &1.09  \\

    \hline
\end{tabular}
\end{center}
\end{table}
%%%%%%%%%%%%%%%%%%%%%%%%%%%%%%%%%%%%%%%%%%%%%%%%%%%%%%%%%%%%%%%%%%%%%%%%%%%%%%%%%%%%%%%%%%%%%%%%%%%%%%%%%%%%%%%%%%%%%%%%%%
%  End of Table 8
%%%%%%%%%%%%%%%%%%%%%%%%%%%%%%%%%%%%%%%%%%%%%%%%%%%%%%%%%%%%%%%%%%%%%%%%%%%%%%%%%%%%%%%%%%%%%%%%%%%%%%%%%%%%%%%%%%%%%%%%%%

%%%%%%%%%%%%%%%%%%%%%%%%%%%%%%%%%%%%%%%%%%%%%%%%%%%%%%%%%%%%%%%%%%%%%%%%%%%%%%%%%%%%%%%%%%%%%%%%%%%%%%%%%%%%%%%%%%%%%%%%%%
%%%%%%%%%%%%%%%%%%%%%%%%%%%%%%%%%%%%%%%%%%%%%%%%%%%%%%%%%%%%%%%%%%%%%%%%%%%%%%%%%%%%%%%%%%%%%%%%%%%%%%%%%%%%%%%%%%%%%%%%%%
% Conclusions
\section{Conclusions}\label{sec:conclusions}
%%%%%%%%%%%%%%%%%%%%%%%%%%%%%%%%%%%%%%%%%%%%%%%%%%%%%%%%%%%%%%%%%%%%%%%%%%%%%%%%%%%%%%%%%%%%%%%%%%%%%%%%%%%%%%%%%%%%%%%%%%
%%%%%%%%%%%%%%%%%%%%%%%%%%%%%%%%%%%%%%%%%%%%%%%%%%%%%%%%%%%%%%%%%%%%%%%%%%%%%%%%%%%%%%%%%%%%%%%%%%%%%%%%%%%%%%%%%%%%%%%%%%
In this article we have studied the ground and excited state properties of the ThO and ThS molecules using the relativistic formulation of the CCSD, CCSD(T), and IHFSCC methods. 
We show that one has to correlate 34 electrons in the reference CCSD state to include the most important electron correlation effects in both systems. 
However, to reach a spectroscopic accuracy and reproduce experimental spectroscopic constants of ThO and ThS, the inclusion of triples correction (CCSD(T)) becomes indispensable.  
Our best prediction for the ground-state equilibrium bond lengths and vibrational frequencies comes from the CCSD(T) approach extrapolated to the basis set limit. 
Specifically, these spectroscopic parameters are 1.842~\AA~and 896 cm$^{-1}$ for ThO, and 2.345~\AA~and 488 cm$^{-1}$ for ThS. 
Neglecting the contributions form triple excitations brings the theoretical vibrational frequencies by approximately 30 cm$^{-1}$ apart from experimental (and CCSD(T)) reference values. 
Moreover, we demonstrated that acceptable spectroscopic constants can be obtained from a triple zeta quality basis set. 

Our study indicates that the spin--orbit electronic spectra of ThO and ThS bear a lot of similarities and are rather different from the ThF$^+$ molecule. 
Specifically, both molecules have the same character and order of excitation energies with the only difference that the first three excited states of ThS appear at somehow lower energy ranges. 
Our spectroscopic constants for the ground and excited states of ThO are in very good agreement with experiment. 
Therefore, we believe that our spectroscopic constants for the ThS molecule will serve as reference values, where experimental spectra in the lowest-lying region are not available. 
Furthermore, we provided new reference values (form the IHFSCC(1,1) method) for the excitation energies of the $^3\Delta_2$ state in both systems. They are 5 168 and 2 616 cm$^{-1}$ for ThO and ThS, respectively. 
These numbers can be further used in the evaluation of the lower bound of the eEDM. 
In addition, we calculated the IP$_1$ and EA$_1$ for both the ThO and ThS molecules.  
The IPs are 6.69 and 6.88 eV, and the EAs are 0.58 and 1.09 eV for ThO and ThS, respectively. 

Finally, we demonstrated that qualitatively correct electronic spectra of ThO and ThS can also be obtained from sector (0,2) of the Fock space without much loss of accuracy.  
Both the main character of electronic transition and their correct order are nicely reproduced compared to the IHFSCC(1,1) method. 
The differences in adiabatic excitation energies amount to at most 2 000 cm$^{-1}$. 
The agreement between sectors (1,1) and (0,2) of the Fock space is better for ThO than for ThS. 
The composition of the wave function character for sectors (1,1) and (0,2) of the Fock space is more similar for ThO than ThS. 
The advantage of the IHFSCC(0,2) approach is, however, its less susceptibility to admixture of undesired $P_i$ contaminated states. 
As a result, in both the ThO and ThS molecules, we obtained a more complete set of excitation energies from sector (0,2) of the Fock space.  

Our study suggest that the application of the IHFSCC(0,2) method for ionically bonded molecules such as ThO and ThS should give qualitatively correct and complete electronic spectra. 
This is particularly important for cases where the IHFSCC(1,1) electronic spectra could not be converged.  

%%%%%%%%%%%%%%%%%%%%%%%%%%%%%%%%%%%%%%%%%%%%%%%%%%%%%%%%%%%%%%%%%%%%%%%%%%%%%%%%%%%%%%%%%%%%%%%%%%%%%%%%%%%%%%%%%%%%%%%%%%
%%%%%%%%%%%%%%%%%%%%%%%%%%%%%%%%%%%%%%%%%%%%%%%%%%%%%%%%%%%%%%%%%%%%%%%%%%%%%%%%%%%%%%%%%%%%%%%%%%%%%%%%%%%%%%%%%%%%%%%%%%
%  Acknowledgment
%%%%%%%%%%%%%%%%%%%%%%%%%%%%%%%%%%%%%%%%%%%%%%%%%%%%%%%%%%%%%%%%%%%%%%%%%%%%%%%%%%%%%%%%%%%%%%%%%%%%%%%%%%%%%%%%%%%%%%%%%%
%%%%%%%%%%%%%%%%%%%%%%%%%%%%%%%%%%%%%%%%%%%%%%%%%%%%%%%%%%%%%%%%%%%%%%%%%%%%%%%%%%%%%%%%%%%%%%%%%%%%%%%%%%%%%%%%%%%%%%%%%%
\section{Acknowledgement}
P.T. thanks the POLONEZ fellowship program of the National Science Center (Poland), No. 2015/19/P/ST4/02480. This project had received funding from the European Union's Horizon 2020 research and innovation programme under the Marie Sk{\l}odowska--Curie grant agreement No. 665778.
C.E.G.E gratefully acknowledges CONACYT and the Secretary of Innovation, Science and Technology of the State of Morelos for the scholarship for graduate  studies, and financial support from the Natural Sciences and Engineering Research Council of Canada. 
Calculations have been carried out using resources provided by Wroclaw Centre for Networking and Supercomputing (http://wcss.pl), grant no.~411. 
The authors gratefully acknowledge computer time provided by the Boguslawski research group at NCU Torun.
%%%%%%%%%%%%%%%%%%%%%%%%%%%%%%%%%%%%%%%%%%%%%%%%%%%%%%%%%%%%%%%%%%%%%%%%%%%%%%%%%%%%%%%%%%%%%%%%%%%%%%%%%%%%%%%%%%%%%%%%%%%%%%%%%%%%%%%%%%%%%%%%%%%
%%%%%%%% REFERENCES      %%%%%%%%%%%%%%%%%%%%%%%%%%%%%%%%%%%%%%%%%%%%%%%%%%%%%%%%%%%%%%%%%%%%%%%%%%%%%%%%%%%%%%%%%%%%%%%%%%%%%%%%%%%%%%%%%%%%%%%%%%
%%%%%%%%%%%%%%%%%%%%%%%%%%%%%%%%%%%%%%%%%%%%%%%%%%%%%%%%%%%%%%%%%%%%%%%%%%%%%%%%%%%%%%%%%%%%%%%%%%%%%%%%%%%%%%%%%%%%%%%%%%%%%%%%%%%%%%%%%%%%%%%%%%%
%\normalem
\bibliography{rsc} 
\bibliographystyle{rsc}
\end{document}